\documentclass[aps,prc,twocolumn,floatfix,nofootinbib,superscriptaddress]{revtex4-1}
\usepackage{float}
\usepackage[T1]{fontenc}
\usepackage[utf8]{inputenc} 
\usepackage{amsmath,amsfonts,amssymb}
\usepackage{graphicx}
\usepackage{epstopdf} 
\usepackage{xcolor}
\usepackage{hyperref} 
\hypersetup{colorlinks=true,linkcolor=blue,citecolor=blue,urlcolor=blue}
\usepackage{bm}

\begin{document}

\title{Bayesian Framework for the $S_{E1}(300\,{\rm keV})$- and $S_{E2}(300\,{\rm keV})$- Factors 
for \texorpdfstring{$\bm{{}^{12}\mathrm{C}(\alpha,\gamma){}^{16}\mathrm{O}}$}{12C(a,g)16O}
from Subthreshold and Ground-State Asymptotic Normalization Coefficients}
\author{A.\ M.\ Mukhamedzhanov}
\affiliation{Texas A\&M University, College Station, TX 77843, USA}

\begin{abstract}
The $^{12}\mathrm{C}(\alpha,\gamma)^{16}\mathrm{O}$ reaction governs the carbon-
to-oxygen ratio set during helium burning, shaping white-dwarf structure and 
Type~Ia supernova yields. At the astrophysical energy $E \approx 300~\mathrm{keV}$, 
the cross section is controlled by the subthreshold $1^{-}$ 
(7.12~MeV) and $2^{+}$ (6.92~MeV) states, whose contributions depend on their 
asymptotic normalization coefficients (ANCs) $C_{1}$ and $C_{2}$, respectively. 
We perform a Bayesian analysis of the $S_{E1}$ and 
$S_{E2}$ factors using calibrated $R$-matrix mappings and 
contemporary ANC constraints for the $1^{-}$, $2^{+}$, and $0^{+}$ ground state. 
For $S_{E1}$, a flat prior on the $1^{-}$ ANC yields a broad 
posterior (MAP $\approx 95.5$~keV\,b), while Gaussian priors centered at the low 
and high anchors produce narrower posteriors with MAPs $\approx 75.4$ and 
$\approx 93.2$~keV\,b, respectively. For $S_{E2}$, we employ a 
calibrated quadratic map in $(C_{2},C_{0})$ that captures the interference 
between the subthreshold $2^{+}$ capture and the external direct capture; the 
resulting posteriors reflect how the joint $(C_{2},C_{0})$ posterior loads the 
interference corridor. The resulting $S(300)$ range is consistent with 
stellar–evolution models that produce low–spin, high–mass ($\sim 
50$–$70\,M_{\odot}$) black holes. Overall, the Bayesian framework yields statistically robust posteriors for 
$S_{E1}$ and $S_{E2}$, quantifying 
correlations and uncertainties relevant for stellar modeling.  
This work links the microscopic subthreshold ANCs ($C_{1},C_{2},C_{0}$) to a
macroscopic astrophysical outcome—the formation of massive, low–spin black 
holes—thereby connecting nuclear reaction experiments and theory with stellar evolution, black–hole demographics, and gravitational–wave observations.
\end{abstract}

\maketitle

\section{Introduction}

The $^{12}\mathrm{C}(\alpha,\gamma)^{16}\mathrm{O}$ reaction is a key process in 
nuclear astrophysics. Together with the triple-$\alpha$ reaction, it determines 
the carbon-to-oxygen ratio established at the end of core helium burning in 
massive and intermediate-mass stars. This ratio influences subsequent stellar 
evolution, the internal composition profiles of white dwarfs, and the 
nucleosynthetic yields of Type Ia supernovae\cite{Iliadis,Woosley2002}. Reliable 
predictions of these outcomes require an accurate determination of the 
$S$-factor at the Gamow-window energy $E \approx 300$~keV, where direct 
measurements are not available due to the exponentially suppressed reaction 
cross section.

The importance of the $^{12}\mathrm{C}(\alpha,\gamma)^{16}\mathrm{O}$ rate 
extends beyond late-stage stellar burning. The final carbon–oxygen core mass of 
massive stars depends sensitively on the ratio of $S_{E1}/ 
S_{E2}$\cite{Farmer2019,Woosley2021}. This core mass, in turn, 
determines whether the star undergoes direct collapse, pulsational 
pair–instability mass ejection, or complete pair–instability disruption. 
Consequently, the $^{12}\mathrm{C}(\alpha,\gamma)^{16}\mathrm{O}$ rate affects 
the maximum mass of black holes formed through stellar collapse and therefore 
influences the location of the lower boundary of the observed “black-hole mass 
gap” inferred from gravitational-wave 
observations\cite{LIGO2021,SperaMapelli2017}. In this sense, the same nuclear physics that controls helium burning composition 
also constrains compact-object demographics.  

Recent gravitational–wave observations of low–spin, high–mass black holes ($\sim 
50$–$70\,M_{\odot}$) \cite{Wang2024,LIGO2021,Wang} have renewed interest in 
refining $S(300)$ to constrain these evolutionary pathways. 
Such objects naturally arise in stellar-evolution models in which the 
$^{12}\mathrm{C}(\alpha,\gamma)^{16}\mathrm{O}$ rate is relatively low, 
$S \approx 110$~keVb, since a reduced $S$-factor diminishes 
carbon consumption during helium burning and therefore increases the final 
carbon–oxygen core mass, suppressing the onset of pulsational pair-instability 
and allowing heavier black-hole remnants to form.

At low energies, the capture amplitude is dominated by the near-threshold 
subthreshold states of $^{16}$O: the $1^{-}$ state at 7.12MeV and the $2^{+}$ 
state at 6.92MeV. Their contributions to the radiative capture arise through the 
long-range tails of their bound-state wave functions in the external region. 
The strength of these contributions is determined by the corresponding 
asymptotic normalization coefficients (ANCs), $C_{1}$ and $C_{2}$, together with 
the ground-state ANC $C_{0}$ controlling the external direct-capture amplitude. 
Thus, accurate determination of the ANCs is central to reducing the 
uncertainties in $S_{E1}$ and 
$S_{E2}$~\cite{muk2024}.

In this work, we perform a Bayesian determination of $S_{E1}(300~\mathrm{keV})$ 
and $S_{E2}(300~\mathrm{keV})$ using calibrated $R$-matrix mappings and 
contemporary ANC constraints, with explicit propagation of parameter 
correlations arising from interference between the subthreshold resonance and 
external direct capture amplitudes and higher resonances. Bayesian statistical methods provide a systematic framework to address the 
remaining uncertainties. Within the $R$-matrix description, Bayesian inference 
combines prior information on ANCs (from radiative captures, transfer reactions, 
elastic scatterings, or theoretical models) with likelihoods based on capture 
and scattering data \footnote{In our Bayesian analysis the likelihood function 
is constructed from observables that the $R$‑matrix model directly 
predicts—radiative‑capture and elastic‑scattering cross sections.  Information 
from transfer reactions enters through the prior distributions on the asymptotic 
normalization coefficients (ANCs), because modeling transfer cross‑section data 
in the likelihood would require additional reaction‑mechanism assumptions.  
Using ANC values extracted from transfer experiments as priors avoids 
double‑counting and keeps the likelihood tied to observables with a uniform 
theoretical treatment.}. This approach yields full posterior probability 
distributions for the ANCs and, consequently, for the extrapolated $S$-factors. 
Unlike traditional fits that return only central values and approximate error 
bars, the Bayesian framework quantifies correlations, propagates uncertainties 
consistently, and provides well-defined credible intervals (CIs) for $S_{E1}$ 
and $S_{E2}$. Such probabilistic treatments have been applied, for example, in 
recent Bayesian $R$-matrix analyses \cite{Odell}. In the present work, we extend 
this approach by explicitly incorporating the existing ANC constraints for the 
$1^{-}$ and $2^{+}$ subthreshold states and the $0^{+}$ ground state of 
$^{16}$O, thereby producing statistically robust posteriors for the 
astrophysical $S$-factors at $300$~keV.

Our results demonstrate that the microscopic subthreshold ANCs 
($C_{1},C_{2},C_{0}$) map directly onto a macroscopic astrophysical 
phenomenon—the formation of massive, low–spin black holes—providing a 
unique connection between nuclear reaction theory, stellar evolution, 
black–hole populations, and gravitational–wave detections.
 
In what follows, we simplify notation by writing $S_{E i} \equiv S_{E 
i}(300~\mathrm{keV}), \quad i=1,2,\,$
so that $S_{E1}$ and $S_{E2}$ denote the astrophysical $S$--factors evaluated at 
$E=300$~keV. Captions, tables and  section titles  retain the explicit $300$~keV label for 
clarity.

\section{State of Knowledge of Subthreshold ($1^{-}$ and $2^{+}$) and Ground-
State ANCs} 

Table~\ref{Table_ANCs} summarizes the most reliable determinations of the ANCs 
for the
${}^{16}{\mathrm O}^{*}\to \alpha + {}^{12}{\mathrm C}$ channels, derived from 
transfer reactions,
elastic scattering extrapolations, and theoretical analyses.

\begin{table*}[t]
\caption{ANC values $C_\ell$ (fm$^{-1/2}$) for 
$^{16}$O$^*(J^\pi)\to \alpha+{}^{12}$C(g.s.). An en dash indicates not reported.
For \cite{Morais}, three $C_0$ values were obtained for three different
potentials.}
\centering
\renewcommand{\arraystretch}{1.15}
\setlength{\tabcolsep}{4pt}
\begin{tabular}{|c|c|c|c|}
\hline
$C_0$ ($0^+$) & $C_2$ ($2^+$) & $C_1$ ($1^-$) & Ref. \\
\hline
--- & $(1.11\pm0.10)\times10^{5}$ & $(2.08\pm0.19)\times10^{14}$ & \cite{Brune} \\
--- & & $1.90 \times10^{14}$ & \cite{Azuma} \\
--- & $(1.22\pm0.06)\times10^{5}$ & $(2.10\pm0.14)\times10^{14}$ & \cite{Avila} \\
--- & $(0.98\pm0.08)\times10^{5}$ & $(1.83\pm0.08)\times10^{14}$ & 
\cite{Hebborn} \\
---& $(1.07\pm0.06)\times10^{5}$ & $(1.85\pm0.07)\times10^{14}$ & 
\cite{Avila,Hebborn} \\
--- & $(1.40\pm0.42)\times10^{5}$ & $(1.87\pm0.32)\times10^{14}$ & 
\cite{Belhout} \\
--- & $(1.44\pm0.26)\times10^{5}$ & $(2.00\pm0.69)\times10^{14}$ & 
\cite{Oulevsir} \\
--- & $(1.54\pm0.18)\times10^{5}$ &  & \cite{Tischhauser} \\
--- & $1.67 \times10^{5}$ & $1.96 \times 10^{14}$ & \cite{Tang} \\
--- & $1.5 \times10^{5}$ & $1.94 \times 10^{14}$ & \cite{Schurmann} \\
--- & $(1.445\pm0.085)\times10^{5}$ & --- & \cite{Sparenberg2004} \\
--- & $(1.10\text{--}1.31)\times10^{5}$ & $(2.21\pm0.07)\times10^{14}$ & 
\cite{Sparen} \\
--- & $1.34\times10^{5}$ & --- & \cite{Descouvemont1987} \\
--- & $1.26\times10^{5}$ & --- & \cite{Dufour} \\
 $(744\pm144)$ & $(1.56 \pm 0.27)\times 10^{5}$ & $(2.72 \pm 1.60) \times 10^{5}$ &
 \cite{Gupta} \\
$58$ & $1.14\times10^{5}$ & $2.08\times10^{14}$  &  \cite{deBoer} \\
--- & $(1.42\pm0.05)\times10^{5}$ & $(2.27\pm0.02)\times10^{14}$ & \cite{Blokh} 
\\
$(337\pm45)$ & --- & --- & \cite{Shen2023} \\
$709$ & --- & --- & \cite{Sayre} \\
$740$ & --- & --- & \cite{Desc} \\
$(637\pm86)$ & --- & --- & \cite{Adhikari} \\
$3390;\ 1230;\ 790$ & --- & --- & \cite{Morais} \\
\hline
\end{tabular}
\smallskip
reported for three different potentials.\par
\label{Table_ANCs}
\end{table*}

Following Ref.~\cite{Hebborn}, the ANCs of Refs.~\cite{Brune,Avila} have been 
renormalized
to $C_{1}=1.83\times 10^{14}$~ fm$^{-1/2}$ and $C_{2}=0.98\times 10^{5}$~ 
fm$^{-1/2}$, which we adopt as the lower
bounds of our prior intervals, while the $R$--matrix anchors are taken from
Ref.~\cite{muk2024}.

\subsection{ANC $C_{1}$}
\label{ANCC1}
The subthreshold $1^{-}$ ANC adopted for this analysis  lies in the range
$C_{1}=(1.83$--$2.27)\times 10^{14}$ fm$^{-1/2}$.
Although this spread appears modest, it significantly impacts the extrapolated
$S_{E1}$.  
The pioneering determinations of $C_{1}$ and $C_{2}$ were made in the sub-
Coulomb
$\alpha$-transfer experiment 
${}^{12}{\mathrm C}({}^{6}{\mathrm Li},d){}^{16}{\mathrm O}^{*}$ ~\cite{Brune},
later confirmed in~\cite{Avila}.  
Historically, the determinations of $C_{1}$ extracted from
sub-Coulomb $\alpha$-transfer reactions
${}^{12}\mathrm{C}({}^{6}\mathrm{Li},d){}^{16}\mathrm{O}^{*}$
in Refs.~\cite{Brune,Avila} relied on the input
$C_{\alpha d}^{2}=5.3\pm0.5$ fm${}^{-1}$ obtained in
Ref.~\cite{Blokhintsev}.  This  ANC was obtained
using the analytic continuation of the solution of a novel energy-dependent 
phase-shift analysis of elastic 
$d-{}^{4}{\rm He}$ scattering to the pole corresponding to the ${}^{6}{\rm Li}$ 
ground state and  by solving three-body $\alpha+p+n$  Faddeev equations.  Direct
 model-independent extrapolation of the $\alpha +d$ elastic scattering phase 
shifts performed in
\cite{blokh2018}  and  variational Monte Carlo calculations \cite{Nollett} 
corroborated the result obtain in \cite{Blokhintsev}. 

Recently, {\it ab initio} no-core shell model with continuum
calculations have revised this to
$C_{\alpha d}^{2}=6.864\pm0.210$ fm${}^{-1}$~\cite{Hebborn},
a $\sim30\%$ increase, which implies a corresponding
$\sim14\%$ reduction in the deduced $C_{1}$ and $C_{2}$ values.
Accordingly, the renormalized ANCs listed in
Table~\ref{Table_ANCs} reflect this updated $C_{\alpha d}$ constraint.
This change does not supersede the earlier experimentally based
extractions, but it is important for interpreting the degree of
agreement (or tension) among the ANC determinations.

Other determinations of the ${}^{16}{\rm O}$ ANC $C_{1}$ include~\cite{Belhout},
 though with larger
uncertainty due to measurements above the Coulomb barrier, and 
model-independent extrapolation of elastic scattering~\cite{Blokh},
later confirmed by Padé extrapolation in Ref.~\cite{Sparen}.  
For Bayesian inference we therefore define three priors:
a flat prior on $[1.83,2.27]\times10^{14}$ fm$^{-1/2}$,
a Gaussian centered at $C_{1}=1.83\times10^{14}$ fm$^{-1/2}$,
and a Gaussian centered at $C_{1}=2.27\times10^{14}$ fm$^{-1/2}$.

Ordinarily, the credibility interval for $S_{E1}$ would reflect the adopted ${}^{6}\mathrm{Li}$ ANC $C_{\alpha d}$, since it enters directly into the inferred $C_{1}$.
But with a \emph{flat prior} on $C_{1}$, the likelihood is less sensitive to the low-$C_{1}$ region (see Eq.~(\ref{JacobianE1})), making the posterior skewed toward higher $C_{1}$.
Because the low-$C_{1}$ tail is excluded in the central interval, the correlation with $C_{\alpha d}$ becomes negligible for the $68\%$ credible interval of the $S_{E1}$ posterior.  But this correlation becomes important for the low-Gaussian prior.

\subsection{ANC $C_{2}$}
\label{ANCC2}

The adopted prior range for $C_{2}$ parallels that used for $C_{1}$.  
Its lower bound is set by the renormalized transfer value constrained by the updated ${}^{6}\mathrm{Li}$ ANC~\cite{Hebborn}, while the upper bound follows from the extrapolation of the $\alpha+{}^{12}\mathrm{C}$ elastic-scattering phase shifts in Ref.~\cite{Blokh}.  
The original sub-Coulomb results of Refs.~\cite{Brune,Avila} lie within these limits, so the chosen prior naturally encompasses all existing determinations.  
Accordingly, three alternative priors are considered for $C_{2}$:  
(i) a flat prior over $C_{2}\in[0.98,\,1.42]\times10^{5}~\mathrm{fm}^{-1/2}$;  
(ii) a Gaussian prior centered at the renormalized transfer value; and  
(iii) a Gaussian prior centered at the higher value inferred from elastic-scattering extrapolation.

As in the case of $C_{1}$, the credible interval associated with $C_{2}$ retains a correlation with the ${}^{6}\mathrm{Li}$ ANC, $C_{\alpha d}$, because the normalization used to infer $C_{2}$ from transfer reactions depends directly on $C_{\alpha d}$.  
However, in contrast to the $E1$ case, the $E2$ amplitude depends on both $C_{2}$ and the ground-state ANC $C_{0}$.  
Therefore, the uncertainty associated with $C_{\alpha d}$ propagates through a two-dimensional $(C_{2},C_{0})$ parameter space rather than through $C_{2}$ alone, shifting the effective prior region in the direction of lower $C_{2}$ values.  
This lower $C_{2}$  prior domain plays an important role in shaping the subsequent $S_{E2}$ posterior once the mapping and likelihood constraints are introduced.  But this correlation becomes especially important for the low-Gaussian prior.

\section{Bayesian formalism}
\subsection{Bayes' theorem and working space}

Our goal is to infer the posterior distributions of the astrophysical 
factors $S_{E1}$ and $S_{E2}$. 
Bayes' theorem for a generic parameter $x$ reads
\begin{align}
P(x\mid D)=\frac{{\mathcal L}(D\mid x)\,P(x)}{P(D)} ,
\label{eq:bayesS}
\end{align}
where  $P(x\mid D)$ is the \emph{posterior}, i.e.\ the probability density of 
$x$ 
given the data $D$;
${\mathcal L}(D\mid x)$ is the \emph{likelihood}, quantifying the probability of
 observing the data $D$ if $x$ is assumed  (see Appendix  \ref{LiPr});
$P(x)$ is the \emph{prior}, encoding information known about $x$ 
    before considering the new data;
 $P(D)$  is the normalization factor, also called the Bayesian evidence. It 
serves only as a normalization constant to ensure that the posterior integrates 
to unity:
$\int P(x\mid D)\,dx = 1$.  
Note that both $P(D)$ and $P(x)$ are marginal probabilities: 
$P(D)$ is the marginal probability of the data (evidence), while 
$P(x)$ is the marginal probability of the parameter space (prior) (see Appendix 
  \ref{app:marginals}) .  

In the current scenario, the primary quantities constrained by experiment are 
the ANCs. 
The astrophysical factors $S_{E1}$ and $S_{E2}$ are derived observables 
that follow from the $R$--matrix mappings $S_{E1}(C_{1})$ and $S_{E2}(C_{2},\, 
C_{0})$. 
Hence the posterior for $S_{Ei}$ ($i=1,2$) is obtained by transforming the 
posterior 
in $C$ through these maps.

\subsection{ANCs as parameters}

Experimental information (from transfer reactions, elastic scattering, and 
theoretical inputs) provides estimates ${\tilde C}$ of the true but unknown 
ANCs. 
These measurements are typically modeled by Gaussian likelihoods,
\begin{align}
P({\tilde C}\mid C) \;\propto\; 
\exp\,\left[-\sum_{k}\frac{({\tilde C}^{(k)}-C)^{2}}{2\sigma_{k}^{2}}\right],
\label{eq:PCtildC}
\end{align}
with uncertainties $\sigma_{k}$. The prior distribution $P(C)$ specifies the 
physically allowed ANC range. In cases where no single value is preferred, 
we adopt a uniform prior across the interval summarized in 
Table~\ref{Table_ANCs}. Combining likelihood and prior yields the ANC posterior:
\begin{align}
P(C\mid D) \;\propto\; {\mathcal L}({\tilde C}\mid C)\,P(C).
\end{align}

\subsection{Transformation to $S$--factors}

Because $S_{Ei}$ is not an independent parameter but a deterministic function 
of the ANC(s), the posterior for $S_{Ei}$ follows by integration over $C$ (also 
called variable change, propagation or push-forward):
\begin{align}
P(S_{Ei}\mid D) &= \int \delta\,\Big(S_{Ei} - S_{Ei}(C)\Big)\,
P(C\mid D)\,dC .
\label{eq:intCtC}
\end{align}
The $\delta$--function ensures that probability weight in ANC space is mapped 
exactly onto the corresponding values of the $S$--factor. Equivalently, when 
the mapping $S_{Ei}(C)$ is monotone on the prior support, this expression 
reduces to
\begin{align}
P(S_{Ei}\mid D) \;=\; 
P\big(C(S_{Ei}) \mid D\big)\;
\left|\frac{dS_{Ei}}{dC}\right|^{-1},
\label{eq:pushforward1}
\end{align}
Here $C(S_{E})$ denotes the \emph{pre--image} of $S_{E}$ under the
calibration map $g(C)=S_{E}$.  In other words, for a given value
of the astrophysical factor $S_{E}$, $C(S_{E})$ is the ANC value
that maps to it through the deterministic relation $S_{E}=g(C)$.
Thus $S_{E}$ is the image of $C$, while $C(S_{E})$ is its
corresponding pre--image. 

Thus the Bayesian workflow is:
(i) specify prior distributions for the ANCs,  
(ii) update with experimental likelihoods to obtain ANC posteriors,  
(iii) propagate these posteriors through the calibrated $R$--matrix mappings to 
obtain full posterior distributions for $S_{E1}$ and $S_{E2}$.

\subsection{Priors and likelihoods (single--energy treatment at $300$~keV)}
\label{PrLhs}

In Bayesian inference, a prior distribution (or simply prior) specifies the 
probability assigned to model parameters before incorporating the current data. 
It encodes existing knowledge, theoretical constraints, or assumptions, and 
serves as the starting point for applying Bayes’ theorem to obtain the 
posterior. 
Here we introduce priors for the three ANCs relevant at $300$~keV: 
the subthreshold $1^{-}$ ANC $C_{1}$, the subthreshold $2^{+}$ ANC $C_{2}$, 
and the ground--state ANC $C_{0}$.

\paragraph{Flat priors.}  
Our analysis of the literature values summarized in Table~\ref{Table_ANCs} 
indicates that $C_{1}$ is constrained to the interval 
$[1.83,\,2.27]\times10^{14}$~fm$^{-1/2}$ and $C_{2}$ to 
$[0.98,\,1.42]\times10^{5}$~fm$^{-1/2}$, without preference for any subregion.  
Accordingly, we adopt independent uniform priors
\begin{align}
p(C_{1}) &= \mathcal{U}[\bigl(1.83,\,2.27\bigr)\times10^{14}~{\mathrm 
fm}^{-1/2}], 
\label{eq:pC1} \\[4pt] 
p(C_{2}) &= \mathcal{U}[\bigl(0.98,\,1.42\bigr)\times10^{5}~{\mathrm 
fm}^{-1/2}], 
\label{eq:pC2}
\end{align}
where $\mathcal{U}[C^{l},C^{h}]$ denotes a uniform distribution on 
$[C^{l},C^{h}]$.  

\paragraph{Gaussian priors.}  
For scenarios where specific ANC determinations are emphasized, we employ 
truncated normal distributions,
\begin{align}
p(C) = \mathrm{N_{T}}[\mu,\sigma;\,C^{l},C^{h}],
\label{eq:pi}
\end{align}
with mean $\mu$, standard deviation $\sigma$, and truncation limits 
$C^{l},C^{h}$ consistent with physical ranges.  Subscript  $T$ stands for the 
truncation.

\paragraph{Likelihoods.}  
At $300$~keV the ANC constraints are inherited from prior determinations in 
the literature (Table~\ref{Table_ANCs}), rather than from new cross--section 
data. Each published ANC value is reported with a central estimate and 
uncertainty, which we represent by Gaussian likelihoods.  
For the $E1$ channel, the likelihood for a candidate $C_{1}$ is
\begin{align}
\mathcal{L}_{E1}(D_{1}\mid C_{1})
= \frac{1}{\sqrt{2\pi}\,\sigma_{1}}\,
   \exp\,\left[-\frac{(C_{1}-\mu_{1})^{2}}{2\sigma_{1}^{2}}\right],
\label{eq:LE1DC}
\end{align}
where $D_{1}$ denotes the experimental determination of the subthreshold 
$1^{-}$ ANC, summarized by a central value $\mu_{1}$ with quoted uncertainty 
$\sigma_{1}$. In other words, $\mathcal{L}_{E1}(D_{1}\mid C_{1})$ quantifies 
the probability of obtaining the measured result $D_{1}$ if the true ANC were 
$C_{1}$.

For the $E2$ channel\footnote{
The likelihood in Eq.~(\ref{eq:LE2DC}) depends only on $C_{2}$ because it
represents the constraint coming from the experiments that determine
the subthreshold $2^{+}$ ANC, such as $\alpha$-transfer measurements or
elastic-scattering phase–shift extrapolations.  
These experimental observables are sensitive only to $C_{2}$ and do not
involve the ground-state ANC $C_{0}$; therefore $C_{0}$ does not appear
in $\mathcal{L}_{E2}(D_{2}\mid C_{2})$.  
The joint dependence on both $C_{0}$ and $C_{2}$ enters only in the
radiative-capture likelihood $\mathcal{L}(D_{\rm cap}\mid C_{0},C_{2})$
used to determine the posterior for $S_{E2}(300\,\mathrm{keV})$.}
\begin{align}
\mathcal{L}_{E2}(D_{2}\mid C_{2})
= \frac{1}{\sqrt{2\pi}\,\sigma_{2}}\,
   \exp\,\left[-\frac{(C_{2}-\mu_{2})^{2}}{2\sigma_{2}^{2}}\right].
\label{eq:LE2DC}
\end{align}
Here $\mu_{i}$ and $\sigma_{i}$ denote the reported central values and 
uncertainties for each ANC.  
The distinction between likelihoods and priors is elaborated in 
Appendix~\ref{LiPr}.  

\subsection{Structure of the $E1$ capture amplitude}
\label{subsec:E1_amplitude}

At low energies the $E1$ capture is governed by the coherent contribution of the
subthreshold $1^{-}$ state at $E_{x}=7.12$~MeV and the higher $1^{-}$ resonances
above threshold, most importantly the broad $1^{-}$ level at $E_{x}=9.585$~MeV,
together with a smooth background. The direct (nonresonant) $E1$ transition to 
the ground state is strongly isospin suppressed and can be neglected. In the 
$R$--matrix formalism, the amplitude of the  $1^{-}$ subthreshold resonance is proportional to the ANC $C_{1}$,
\begin{equation}
\mathcal{M}_{\text{sub}}(E)\propto C_{1}.
\end{equation}

The total $E1$ amplitude arises from the two-level $R$--matrix interference 
between this subthreshold state and the broad $1^{-}$ resonance with small addition from
 the background resonance. This interference determines
both the absolute normalization and the sign of the energy dependence. The 
resulting astrophysical factor takes the quadratic form
\begin{align}
S_{E1}(E) = \alpha(E)\,C_{1}^{2}+\beta(E)\,C_{1}+\gamma(E),
\label{eq:SE1_quad}
\end{align}
where the $C_{1}^{2}$ term represents the squared contribution of the 
subthreshold
state, the linear term reflects its interference with the $1^{-}$ resonances 
above
threshold, and the constant term arises entirely from the higher $1^{-}$ sector.

At $E=300$~keV the capture is dominated by  two states: the combined
contribution of the subthreshold $1^{-}$ level and the broad $1^{-}$ resonance
accounts for about $95\%$ of the total $S_{E1}$. The remaining
few-percent correction originates from higher $1^{-}$ resonances and the smooth 
background. Moreover, at this
energy the interference between the subthreshold and broad $1^{-}$ amplitudes is
\emph{constructive}, i.e., the interference term 
$\beta\,C_{1}$ is
positive and increases the resulting $S_{E1}$ relative to the subthreshold 
contribution
alone.

\section{Deterministic map for $E1$}
\label{sec:E1_map}

Our Bayesian analysis uses the calibrated $R$--matrix mapping of
Ref.~\cite{muk2024}, which connects the ANC $C_{1}$ of the subthreshold $1^{-}$
state to the astrophysical factor $S_{E1}$.  
Specializing Eq.~\eqref{eq:SE1_quad} to $E=300$~keV yields the calibrated map
\begin{equation}
S_{E1}= s_{1} + \tau_{1}\,C_{1} + \beta_{1}\,C_{1}^{2},
\label{eq:E1map_exact}
\end{equation}
with coefficients determined from the three $R$--matrix anchor points:
\begin{align}
s_{1}   &= -147.23\;\;\text{keV\,b}, \nonumber\\
\tau_{1}&= \;\,1.51\times10^{-12}\;\;\text{keV\,b fm}^{1/2}, \nonumber\\
\beta_{1}&= -1.90\times10^{-27}\;\;\text{keV\,b fm}.
\end{align}

Equation~\eqref{eq:E1map_exact} reproduces exactly
\begin{align}
S_{E1}=68.42~\text{keV\,b} &\quad \text{at}\quad C_{1}=1.83\times10^{14}\ 
\mathrm{fm^{-1/2}},\nonumber\\
S_{E1}=85.0~\text{keV\,b}  &\quad \text{at}\quad C_{1}=2.08\times10^{14}\ 
\mathrm{fm^{-1/2}},\nonumber\\
S_{E1}=98.0~\text{keV\,b}  &\quad \text{at}\quad C_{1}=2.27\times10^{14}\ 
\mathrm{fm^{-1/2}}.\nonumber
\end{align}
This map is fixed by reaction dynamics and is independent of prior choice.
Priors on $C_{1}$ simply determine how probability weight is distributed along
Eq.~\eqref{eq:E1map_exact}, producing either broad (flat prior) or narrow
(Gaussian prior) posteriors for $S_{E1}$.
Flat priors spread weight evenly 
across the full ANC range, yielding broad posteriors skewed toward higher 
$S_{E1}$ due to the negative $\beta_{1}$. Gaussian priors, in contrast, 
concentrate weight near specific ANC values (low or high anchors), producing 
narrower, more localized posteriors.

\subsection{Taylor expansion}
\label{sec:Taylor}

It is worth noting that the quadratic dependence on the ANC employed in 
Eq.~(\ref{eq:E1map_exact}) is mathematically equivalent to a Taylor expansion of
 the corresponding $S$-factor around the 
reference ANCs from Ref.~\cite{deBoer}, e.g.
\begin{align}
& S_{E1}(C_1) \approx S_{E1}(C_{1}^{(0)}) + 
\left.\frac{\partial S_{E1}}{\partial C_1}\right|_{C_{1}^{(0)}} (C_1 - 
C_{1}^{(0)})
\nonumber\\
&+ \frac{1}{2}\left.\frac{\partial^2 S_{E1}}{\partial 
C_1^2}\right|_{C_{1}^{(0)}} (C_1 - C_{1}^{(0)})^2,
\label{eq:Taylor}
\end{align}
and similarly for $S_{E2}(C_2)$. 
The form adopted here in Eq.~(\ref{eq:E1map_exact}) retains the physically 
motivated proportionality $ \propto C_{1}^2$ for the capture through the subthreshold resonance  and  ensures a positive-definite $S$ factor over the full ANC ranges considered in the present Bayesian analysis.

\subsection{Flat prior in $\mathbf{C_{1}}$ and analytic transformation for 
$\mathbf{S_{E1}}$}
For the current scenario we assign a uniform prior to $C_{1}$ on the accepted 
interval:
\begin{equation}
P(C_{1}) \;=\; \frac{1}{C_{1}^{h}-C_{1}^{l}},
\qquad C_{1}\in[C_{1}^{l},\,C_{1}^{h}].
\end{equation}

\subsection{Gaussian prior in $C_{1}$ and transformation to $S_{E1}$}

In addition to the flat prior, we also consider a Gaussian prior for the ANC 
$C_{1}$,  
centered at an anchor value $C_{1}^{\text{anchor}}$ with variance $\sigma^{2}$,
and truncated to the physical interval $[C_{1}^{l},C_{1}^{h}]$:
\begin{equation}
P(C_{1}) \;\propto\; 
\exp\,\left[-\frac{(C_{1}-C_{1}^{\text{anchor}})^{2}}{2\sigma^{2}}\right],
\qquad C_{1}\in[C_{1}^{l},C_{1}^{h}].
\label{eq:Gaussian_prior_C1}
\end{equation}

\subsection{Bayesian transformation and interpretation}
\label{subsec:SE1trans}

The posterior density of $S_{E1}$ is obtained by transforming the prior 
distribution of $C_{1}$ through the exact quadratic map 
(\ref{eq:E1map_exact}).  

\paragraph{Flat prior.}
For a flat prior, $P(C_{1})=1/(C_{1}^{h}-C_{1}^{l})$ on $[C_{1}^{l},C_{1}^{h}]$,
 the posterior is
\begin{align}
P(S_{E1})
&= \frac{1}{C_{1}^{h}-C_{1}^{l}}\,\int_{C_{1}^{l}}^{C_{1}^{h}} 
\delta\,\big(S_{E1} - [s_{1}+\tau_{1}C_{1}+\beta_{1}C_{1}^{2}]\big)\,dC_{1} 
\nonumber\\[6pt]
&= \frac{1}{C_{1}^{h}-C_{1}^{l}}
\sum_{i}\frac{1}{\big|\tau_{1}+2\beta_{1}C_{1}\big|_{C_{1}=C_{1,i}(S_{E1})}},
\label{eq:push_forward_SE1_exact}
\end{align}
where the sum runs over real roots $C_{1,i}(S_{E1})\in[C_{1}^{l},C_{1}^{h}]$ of
$S_{E1}=s_{1}+\tau_{1}C_{1}+\beta_{1}C_{1}^{2}$.

\paragraph{Gaussian prior.}
For the Gaussian prior in Eq.~\eqref{eq:Gaussian_prior_C1}, the posterior is
\begin{align}
P(S_{E1})
&= \frac{1}{Z}\,\int_{C_{1}^{l}}^{C_{1}^{h}}
\exp\,\left[-\frac{(C_{1}-C_{1}^{\text{anchor}})^{2}}{2\sigma^{2}}\right]     
\nonumber\\
&\times \delta\,\big(S_{E1} - 
[s_{1}+\tau_{1}C_{1}+\beta_{1}C_{1}^{2}]\big)\,dC_{1} \nonumber\\[6pt]
&= \sum_{i}
\frac{1}{Z}\,
\frac{\exp\,\left[-\tfrac{(C_{1,i}(S_{E1})-
C_{1}^{\text{anchor}})^{2}}{2\sigma^{2}}\right]}
{\big|\tau_{1}+2\beta_{1}C_{1}\big|_{C_{1}=C_{1,i}(S_{E1})}},
\label{eq:push_forward_SE1_Gauss_exact}
\end{align}
where $Z$ is the normalization constant ensuring $\int P(S_{E1})\,dS_{E1}=1$.

\medskip
\subsection{Interpretation of delta-function representation }  
The delta--function representation clarifies how probability mass in 
$C_{1}$--space maps into $S_{E1}$--space. When the mapping $S_{E1}=g(C_{1})$ 
is monotone, the change of variables gives
\begin{equation}
P(S_{E1}\mid D) \;=\; \frac{P(C_{1}(S_{E1})\mid D)}
{\bigl|\,dg/dC_{1}\,\bigr|_{C_{1}=C_{1}(S_{E1})}},
\end{equation}
where $C_{1}(S_{E1})$ denotes the preimage of $S_{E1}$ under the map $g$.  

The factor $P(C_{1}(S_{E1})\mid D)$ encodes the prior weight at the 
corresponding ANC, while the denominator 
$\lvert dg/dC_{1}\rvert$ quantifies how  $S_{E1}$ responds to 
changes in $C_{1}$. A steep slope ($|dg/dC_{1}|$ large) means that small changes 
 in $C_{1}$ 
produce large changes in $S_{E1}$. In the change-of-variables formula 
the Jacobian factor $|dg/dC_{1}|^{-1}$ then becomes small, so the 
probability density is diluted over a wide range of $S_{E1}$ values to satisfy 
normalization of the posterior (probability density) to unity. 
Conversely, a flat slope ($|dg/dC_{1}|$ small) implies that even large 
changes in $C_{1}$ produce only small variations in $S_{E1}$. In this 
case the Jacobian factor $|dg/dC_{1}|^{-1}$ is large, which concentrates 
probability weight into a narrow region of $S_{E1}$. Thus the shape of 
the posterior in $S_{E1}$ space is controlled not only by the prior in 
$C_{1}$ but also by how sensitively $S_{E1}$ responds to changes in $C_{1}$.  

For the quadratic calibration map
\begin{equation}
g(C_{1}) = s_{1}+\tau_{1}C_{1}+\beta_{1}C_{1}^{2},
\end{equation}
the derivative
\begin{equation}
\frac{dg}{dC_{1}} = \tau_{1}+2\beta_{1}C_{1}
\label{JacobianE1}
\end{equation}
decreases with $C_{1}$ since $\beta_{1}<0$. Thus, the Jacobian factor 
$1/|dg/dC_{1}|$ grows toward larger $C_{1}$, enhancing posterior density 
at higher $S_{E1}$.  

\subsection{Posteriors  for flat and Gaussian priors}
\label{PosterSE1}

\begin{figure}[H]
\centering
\includegraphics[width=0.95\linewidth]{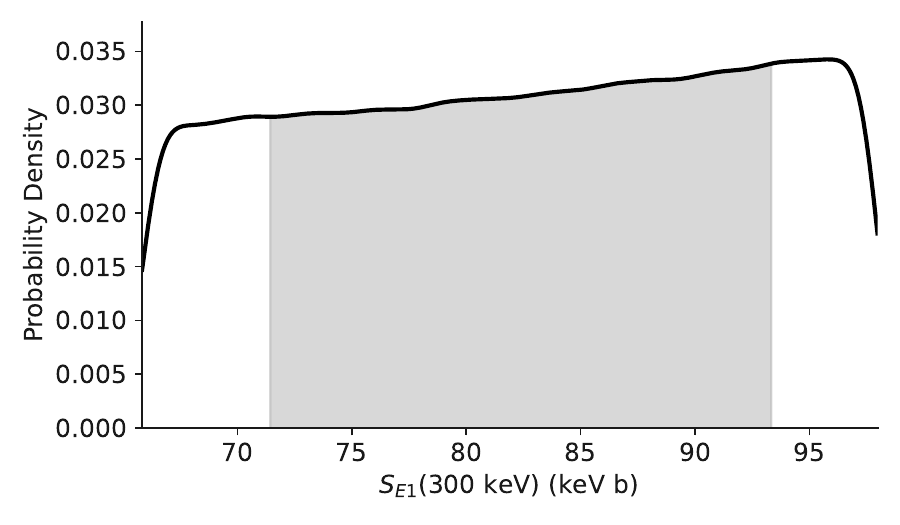}
\caption{Posterior of the $E1$ astrophysical factor obtained by 
propagating a flat prior in the ANC $C_{1}$ through the exact calibration 
map~\eqref{eq:E1map_exact}. Because the mapping is monotone with 
$\beta_{1}<0$, the Jacobian factor $|dS_{E1}/dC_{1}|^{-1}$ enhances 
probability density at larger $C_{1}$, resulting in a right--skewed posterior.
The grey band shows the 68\% Bayesian sredible interval.}
\label{fig:SEflatprior}
\end{figure}

\begin{figure}[H]
\centering
\includegraphics[width=0.95\linewidth]{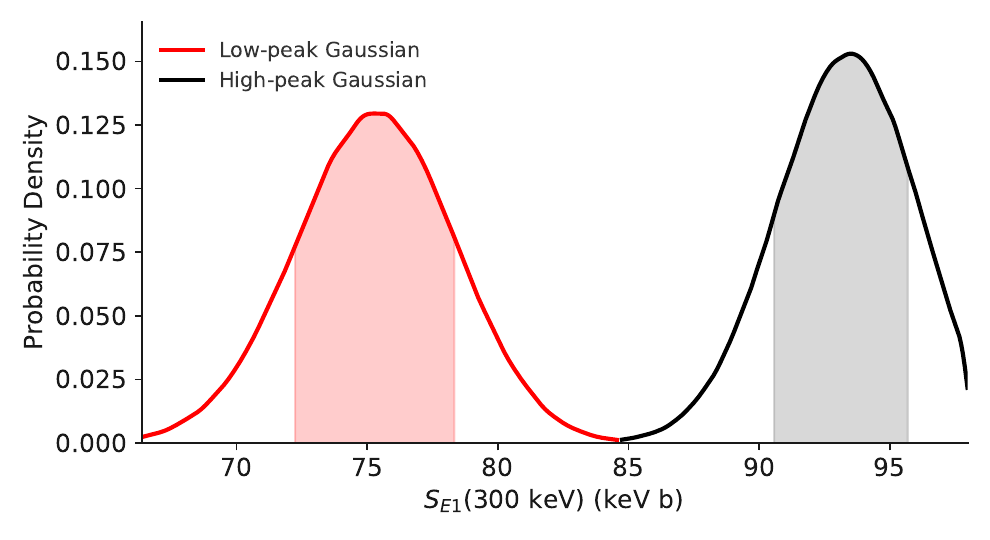}
\caption{Posterior distributions of the $E1$ astrophysical factor 
at $E=300$~keV for Gaussian priors in $C_{1}$.
The red curve corresponds to a low--Gaussian prior centered at the  
$C_{1}=1.83\times10^{14}$~fm$^{-1/2}$, 
while the black curve corresponds to a high--Gaussian prior centered at 
$C_{1}=2.20\times10^{14}$~fm$^{-1/2}$.
Shaded bands indicate 68\% Bayesian credible intervals.}
\label{fig:SEGaussprior}
\end{figure}

The $S_{E1}$ posteriors for  flat and Gaussian priors  are depicted in Figs.  
\ref{fig:SEflatprior}  and 
\ref{fig:SEGaussprior}.  
Flat priors in $C_{1}$ produce right--skewed posteriors in 
$S_{E1}$.  For Gaussian priors in $C_{1}$, the outcome reflects a balance 
between prior
weight and the Jacobian. A high--Gaussian (centered near $C_{1}^{h}$) produces
a posterior strongly peaked at large $S_{E1}$, with the skew reinforced by the 
Jacobian.
A low--Gaussian (centered near $C_{1}^{l}$) yields a posterior around the lower 
anchor,
but the decreasing slope again shifts weight toward the right tail.  
In all cases, the right--skew of the $S_{E1}$ posterior is a direct consequence 
of the
concavity ($\beta_{1}<0$) of the calibration map.

\begin{table*}[t]
\centering
\caption{Posterior summary for $S_{E1}$ using the exact calibrated map
(\ref{eq:E1map_exact}). Entries list the maximum a posteriori (MAP),
the posterior median, and central 68\%/95\% credible intervals (CIs). Priors on 
$C_{1}$ are truncated to
$[1.83,\,2.27]\times 10^{14}~\mathrm{fm}^{-1/2}$.}
\label{tab:SE1_exact_summary}
\begin{tabular}{|l|c|c|c|c|}
\hline
Prior on $C_{1}$ & MAP (keV\,b) & Median (keV\,b) & $68\%$ CI (keV\,b) & $95\%$
   CI (keV\,b) \\
\hline
Flat (uniform)      & 95.5 & 82.8 & [71.4, 93.4] & [66.7, 97.3] \\
Low--Gaussian       & 75.4 & 75.3 & [72.2, 78.3] & [69.2, 81.2] \\
High--Gaussian      & 93.2 & 93.2 & [90.6, 95.7] & [87.9, 97.4] \\
\hline
\end{tabular}
\end{table*}

Note that for the flat prior in $C_{1}$ the posterior is strongly right-skewed:
the MAP (see Appendix \ref{MAPMED})  ($\sim$96~keV\,b) lies outside the central 
$68\%$  credible interval because skewness displaces the posterior from the median.  
In skewed distributions, central credible intervals exclude equal tails of probability
mass and need not contain the MAP.  Because the low-end  tail is excluded, the correlation with ${}^{6}{\rm Li}$  ANC $C_{\alpha\,d}$  becomes not significant.

However, the low–Gaussian posterior remains sensitive to the lower end of the 
$C_{1}$ range, which is correlated with the ${}^{6}\mathrm{Li}$ ANC used to 
normalize the sub–Coulomb transfer extractions \cite{Brune,Avila,Hebborn}.

\section{Analytic Mapping and Bayesian Posterior for $S_{E2}(300~\mathrm{keV})$}
\label{sec:SE2mapping}

The astrophysical $S_{E2}$ factor at $E=300$~keV depends primarily on two 
quantities:
the ANC $C_{2}$ of the subthreshold $2^{+}$ state at $\varepsilon=0.245$~MeV and
the ground--state ANC $C_{0}$ associated with the direct radiative capture to 
the ground state.
Their combined influence can be represented by the calibrated quadratic form
\begin{equation}
S_{E2}(300)
 = s_0 + \alpha\,C_2 + \beta_2\,C_2^2 + \tau_2\,C_2 \,C_0 + \kappa\,C_0^2,
\label{eq:SE2mapping}
\end{equation}
The parameters $(s_0,\alpha,\beta_2,\tau_2,\kappa)$ were determined
by requiring the mapping to reproduce the four physical anchor points
extracted from the $R$-matrix calculations within 
$0.004~\mathrm{keV\,b}$ accuracy.
The quadratic term $C_{2}^{2}$ represents the dominant contribution
from the subthreshold $2^{+}$ resonance.
The inclusion of the linear $C_{2}$ term accounts for the interference
between the subthreshold and higher $2^{+}$ resonances,
among which the level at $E_R=4.35$~MeV provides the principal effect.
The bilinear term $C_{2}C_{0}$ arises from the interference between
the subthreshold resonance and the direct capture to the ground state.
The positive coefficient $\kappa$ quantifies the quadratic increase of 
$S_{E2}$ at large $C_{0}$ values.
A possible linear term in $C_{0}$, associated with the interference 
between the direct-capture amplitude and the higher $2^{+}$ resonances,
was found to be negligible.
 
 \begin{table}[t]
    \centering
    \caption{%
        Calibrated coefficients of Eq.~(\ref{eq:SE2mapping}).
        Units: $S_{E2}$ in keV\,b; $C_0,\,C_2$ in fm$^{-1/2}$.
        The coefficients reproduce the four anchor points
        and ensure a positive $\kappa>0$.}
    \label{tab:SE2coeffs}
    \begin{ruledtabular}
    \begin{tabular}{lcc}
        Coefficient & Numerical value & Unit \\ \hline
        $s_0$   & $-1.9424863844$                   & keV\,b \\
        $\alpha$ & $3.18\times10^{-5}$               & keV\,b$\cdot$fm$^{1/2}$ \\
        $\beta_2$ & $3.6306140021\times10^{-9}$       & keV\,b$\cdot$fm \\
        $\tau_2$ & $-5.1546989137\times10^{-7}$      & keV\,b$\cdot$fm \\
        $\kappa$ & $+1.8464249429\times10^{-5}$      & keV\,b$\cdot$fm \\
    \end{tabular}
    \end{ruledtabular}
\end{table}

\subsection{Three--dimensional $S_{E2}$ surface}
\label{sec:SE2_surface}
\begin{figure}[t!]
    \centering
    \includegraphics[width=0.48\textwidth]{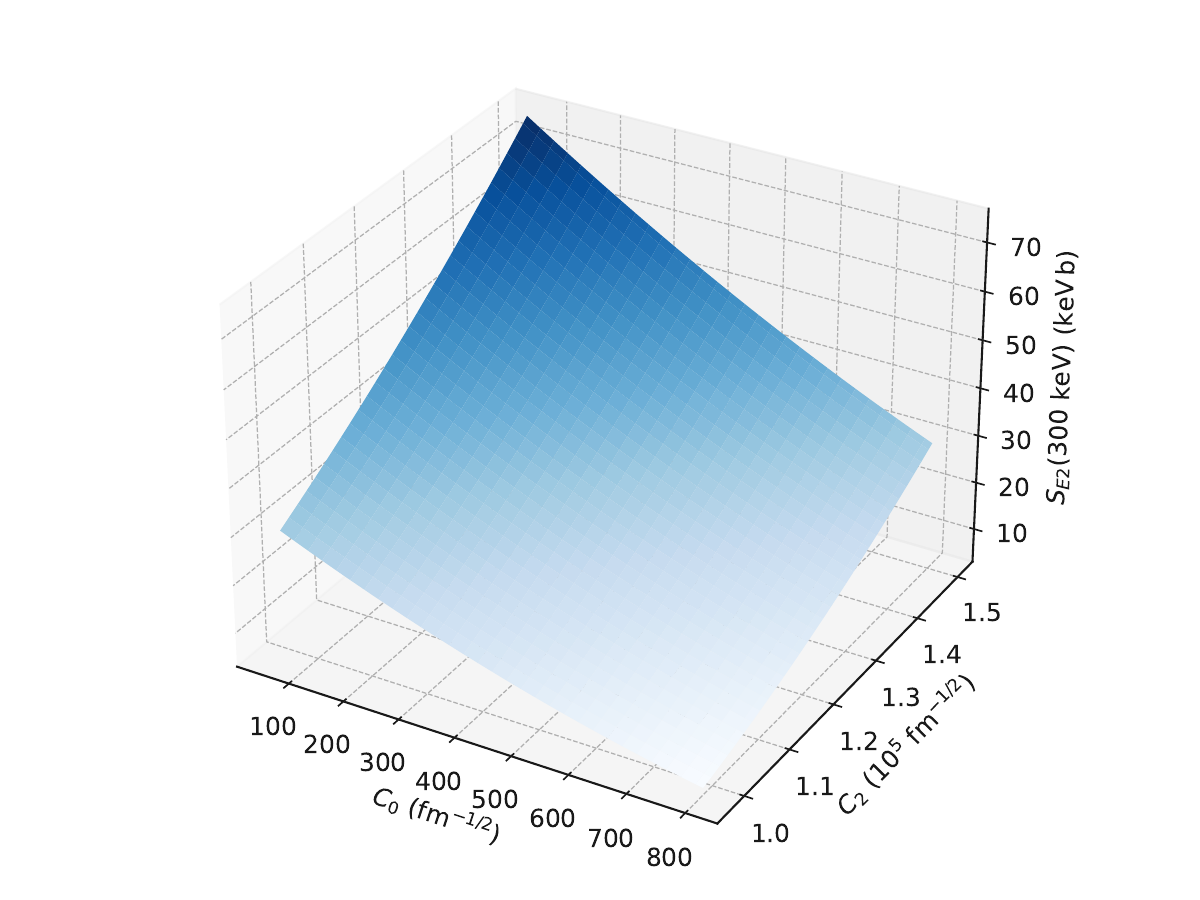}
    \caption{Three--dimensional surface of $S_{E2}(300~\mathrm{keV})$ as a function  of the subthreshold ANC $C_{2}$ and the ground--state ANC $C_{0}$,
        computed using the calibrated mapping of Eq.~(\ref{eq:SE2mapping}). 
        The surface reproduces the $R$-matrix anchor points with better than 
        $0.004~\mathrm{keV\,b}$ accuracy.  
        The steep increase with $C_{2}$ reflects the dominant 
        radiative capture through the subthreshold $2^{+}$ state, while the 
        interference between the subthreshold resonance  and the external nonresonant capture to the ground state generates the bilinear $C_{2}C_{0}$ dependence  responsible for the downward slope with increasing $C_{0}$.  
The positive quadratic term in $C_{0}$ leads to the mild upturn at  large $C_{0}$.  }
  \label{fig:SE2_surface}
\end{figure}

To illustrate the parameter dependence we evaluate the surface
$S_{E2}(C_{2},C_{0})$ over the prior domains
\begin{align}
& p(C_{2}) = \mathcal{U}[\bigl(0.98,\,1.42\bigr)\times10^{5}~{\mathrm 
fm}^{-1/2}],        \nonumber\\
&p(C_{0}) = \mathcal{U}[\bigl(58,\,800~{\mathrm fm}^{-1/2}]    
 \label{pC2C0} 
 \end{align}   
using the calibrated map \eqref{eq:SE2mapping}. This representation reproduces the $R$-matrix anchor points while remaining well behaved across the full parameter space.

Figure~\ref{fig:SE2_surface} shows the three–dimensional surface on a
uniform $(C_{2},C_{0})$ grid with shading in keV\,b.
Two qualitative features are evident:
(i) a steep, nearly quadratic rise with $C_{2}$ from the $\beta_{2}C_{2}^{2}$ 
term,
and (ii) a systematic modulation with $C_{0}$ produced by the interference
term $\tau_{2}C_{2}C_{0}$ together with the quadratic $+\kappa C_{0}^{2}$ term.

\subsubsection{Relative sensitivity to $C_{2}$ and $C_{0}$.}
From Eq.~\eqref{eq:SE2mapping},
\begin{align}
\frac{\partial S_{E2}}{\partial C_{2}} &= \alpha + 2\beta_{2}\,C_{2} + 
\tau_{2}\,C_{0},\\
\frac{\partial S_{E2}}{\partial C_{0}} &= \tau_{2}\,C_{2} + 2\kappa\,C_{0}.
\end{align}
Locally, $\lvert\partial S_{E2}/\partial C_{0}\rvert$ can be comparable to
or exceed $\lvert\partial S_{E2}/\partial C_{2}\rvert$ depending on 
$(C_{2},C_{0})$.
However, the prior ranges differ substantially:
$C_{2}$ spans $\sim 5.0\times10^{4}~\mathrm{fm}^{-1/2}$ whereas
$C_{0}$ spans only $\sim 700~\mathrm{fm}^{-1/2}$.

\subsection{Push-forward structure of the $S_{E2}$ posterior}
\label{Push-forwardSE2}
The posterior for the astrophysical factor at $300$~keV,
$P\,\left(S_{E2}\,\mid\,D\right)$, is obtained as the push-forward (integral 
transformation) of the joint
posterior $P(C_{2},C_{0}\,\mid\,D)$ under the nonlinear map
\begin{align}
S_{E2} \equiv \mathcal{F}(C_{2},C_{0}).
\end{align}
Formally,
\begin{align}
P(s\,\mid\,D)
&= \iint P(C_{2},C_{0}\,\mid\,D)\;\delta\,\bigl(s-
\mathcal{F}(C_{2},C_{0})\bigr)\,dC_{2}\,dC_{0} \nonumber\\
&= \int_{\mathcal{F}(C_{2},C_{0})=s} 
\frac{P(C_{2},C_{0}\,\mid\,D)}{\,\nabla\mathcal{F}(C_{2},C_{0})\,}\;d\ell,
\label{eq:coarea}
\end{align}
where the second line comes from the rule that tells us how probabilities change when we switch from one variable to another (the change-of-variables formula).  The integral is along the level set $\{(C_{2},C_{0}):\mathcal{F}=s\}$, $d\ell$ is 
arclength, and 
\begin{align}
\,\nabla\mathcal{F}\,=\sqrt{\bigl(\partial\mathcal{F}/\partial 
C_{2}\bigr)^{2}+
\bigl(\partial\mathcal{F}/\partial C_{0}\bigr)^{2}}.
\label{JacobianSE2}
\end{align}
$P(C_{2},C_{0}\,\mid\,D)$ denotes the joint posterior probability density for 
the
ANCs $C_{2}$ and $C_{0}$ given the experimental data $D$.  It represents how
strongly the data constrain each parameter \emph{together}, including their
correlation arising from interference between the subthreshold resonance $2^{+}$ and external direct-capture amplitudes.

Equation~(\ref{eq:coarea}) shows explicitly that the push-forward density is 
controlled
by two ingredients: (i) the weight $P(C_{2},C_{0}\,\mid\,D)$ along the level set
$\mathcal{F}=s$, and (ii) the Jacobian factor $1/\,\nabla\mathcal{F}\,\,$.  
Regions in
$(C_{2},C_{0})$ where $\mathcal{F}$ varies \emph{slowly} (small gradient 
magnitude)
contribute most strongly to $P(s\,\mid\,D)$, while regions where $\mathcal{F}$ 
varies
\emph{rapidly} (large gradient) contribute little.  Hence, along the 
interference
corridor  where the map $\mathcal{F}(C_{2},C_{0})$ changes slowly,
$\,\nabla\mathcal{F}\,$ is small and the integrand is enhanced; motion
\emph{perpendicular} to this corridor produces large $\,\nabla\mathcal{F}\,$ and
 is
therefore suppressed.

In practice, the push–forward is evaluated through Monte Carlo sampling of the
joint posterior $P(C_{2},C_{0}\,\mid\,D)$ on the $(C_{2},C_{0})$ grid.  Each
sample is mapped to $S_{E2}$ through 
$\mathcal{F}(C_{2},C_{0})$,
and the resulting distribution of sampled values provides an accurate numerical
representation of $P(S_{E2}\,\mid\,D)$.

To visualize the constraint geometry in $(C_{2},C_{0})$ space, we overlay the
interference ridge (the locus along which $S_{E2}$ varies 
slowly) on the $S_{E2}$ map together with adjacent contour bands. The ridge is 
identified as the diagonal path from lower-left to upper-right in $(C_{2},C_{0})$ where motion along the path preserves the interference balance, while motion across the path leads to rapid changes in $S_{E2}$. 
 \begin{figure}[t!]
 \centering
 \includegraphics[width=0.48\textwidth]{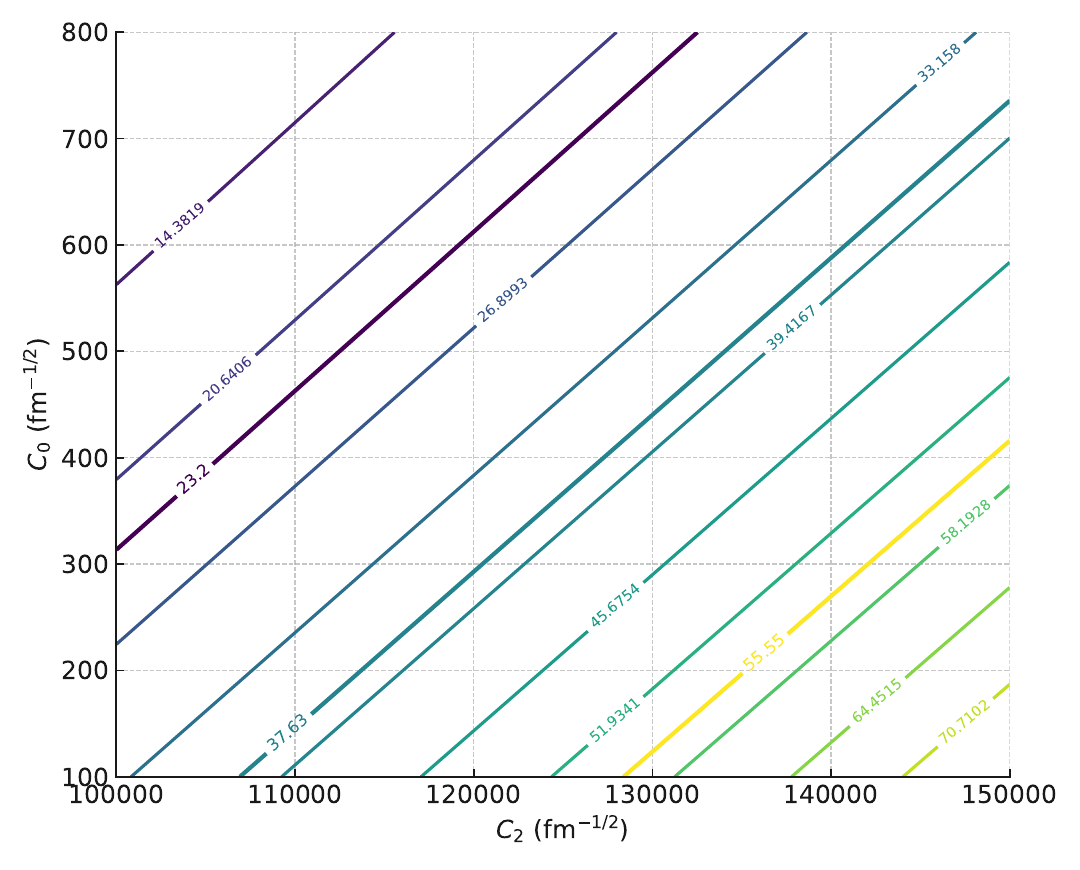}
 \caption{$S_{E2}(300~\mathrm{keV})$ as a function of $(C_{2},C_{0})$.  
The diagonal interference corridor marks the combinations of ANCs for which the
 subthreshold resonance $2^{+}$ contribution and the external direct-capture amplitude interfere in a manner that keeps $S_{E2}$ nearly constant.  Motion across this 
corridor
leads to rapid changes in $S_{E2}$.}
\label{fig:SE2_grid}
\end{figure}

This gives us two main takeaways.  First, the interference corridor spans the entire
$(C_{2},C_{0})$ domain; therefore, the shape of $P(S_{E2}\,\mid\,D)$ is 
determined
by how $P(C_{2},C_{0}\,\mid\,D)$ is distributed \emph{along} this corridor.  
Regions
where $\mathcal{F}(C_{2},C_{0})$ varies slowly dominate the push–forward, while
transverse directions of steep variation are suppressed—precisely in the sense 
of a
saddle–point evaluation.
Second, prior information that restricts $(C_{2},C_{0})$ to particular segments 
of the
corridor modifies the length of the contributing level-set region and the 
Jacobian factor
$1/\|\nabla\mathcal{F}\|$, which narrows $P(S_{E2}\,\mid\,D)$ and suppresses its 
 tails without altering the underlying interference mechanism itself.

\section{Posteriors for quasiflat and Gaussian priors}
\label{Priors}
\subsection{Flat on $C_{2}$ and weighted on $C_{0}$ prior}
\label{quasiflat}

For the Bayesian inference, we adopt a flat prior on the subthreshold
ANC $C_{2}$ within
\begin{align}
C_{2}\in[(0.98,\,1.42)\times10^{5}]~\mathrm{fm^{-1/2}},
\label{C2prior}
\end{align}
and a weighted prior on the ground-state ANC $C_{0}$ over
\begin{align}
C_{0}\in[58,\,800]~\mathrm{fm^{-1/2}}.
\label{C0prior}
\end{align}
The weighted prior on $C_{0}$ is flat for the interval  $[58,\,337]~\mathrm{fm^{-1/2}}$ and  increases linearly from unity at $C_{0}=337~\mathrm{fm^{-1/2}}$.
to a factor of three at $C_{0}=800~\mathrm{fm^{-1/2}}$, consistent with recent theoretical and experimental studies that favor higher ground-state ANCs (see
Table~\ref{Table_ANCs}).  

\begin{figure}[H]
    \centering
    \includegraphics[width=0.48\textwidth]{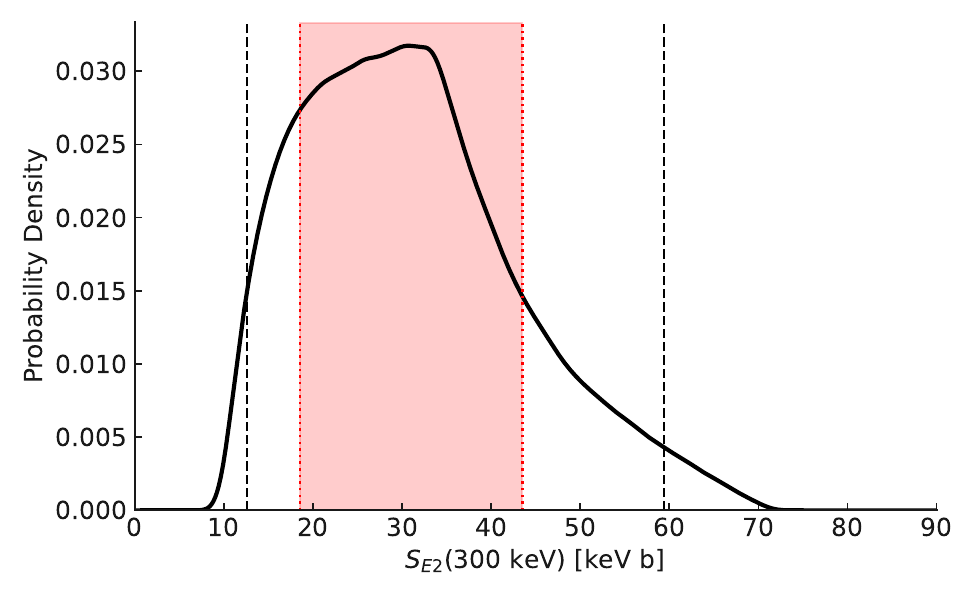}
    \caption{
        Posterior probability density for
        $S_{E2}(300~\mathrm{keV})$
        obtained from Eq.~(\ref{eq:SE2mapping}) using the prior defined
        in Eqs. \ref{C2prior} and \ref{C0prior}.
        The red band denotes the 68\% credible interval and the dashed
        lines mark the 95\% limits.
        The MAP value is
        $S_{E2}(300)=30.6~\mathrm{keV\,b}$,
        with a median of $29.9~\mathrm{keV\,b}$.
        The 68\% and 95\% credible intervals are
        $[18.5,\,43.5]$ and $[12.6,\,59.4]~\mathrm{keV\,b}$,
        respectively.}
    \label{fig:SE2posterior_weightedC0}
\end{figure}

The posterior distribution in Fig.~\ref{fig:SE2posterior_weightedC0}
was obtained through Monte Carlo sampling of $(C_0,\,C_2)$ pairs
according to the weighted priors described above,
and propagated through the calibrated analytic mapping.
The resulting distribution is mildly right-skewed, 
with the MAP  $30.6~\mathrm{keV\,b}$.
This behavior reflects the dominance of the subthreshold $2^+$ capture amplitude
  and the gradual destructive interference with the external component
at large $C_0$.
Estimates from Eq.~(\ref{JacobianSE2}) show that the posterior weight in the low-$C_{2}$ region is enhanced because $\mathcal{F}(C_{2},C_{0})$ decreases when $C_{2}$ becomes small and $C_{0}$ becomes large.  As a result, the $S_{E2}$ posterior is shifted toward lower values, which are dominantly contributed by configurations with small $C_{2}$.  In this regime the correlation between $C_{2}$ and the ${}^{6}\mathrm{Li}$ ANC $C_{\alpha d}$ becomes especially important: the renormalization and uncertainty of $C_{\alpha d}$ directly control how far the posterior can extend into the low-$C_{2}$ region.  This behavior contrasts with the $C_{1}$ case, where the impact of the ${}^{6}\mathrm{Li}$ normalization on the inferred $S_{E1}$ is considerably weaker.

\subsection{Bimodal Gaussian prior on $C_{2}$}
\label{BimodalGaussian}

An alternative formulation employs a bimodal Gaussian mixture as the
prior on the subthreshold ANC $C_{2}$,
representing two physically motivated $R$-matrix solutions.
The first, or ``low-$C_{2}$'',  component is centered at
\begin{align}
C_{2}^{l}=(1.10\pm0.05)\times10^{5}~\mathrm{fm^{-1/2}},
\end{align}
consistent with single-level fits in which the subthreshold
resonance dominates the capture amplitude.
The second, or ``high-$C_{2}$'',  component peaks at
\begin{align}
C_{2}^{h}=(1.42\pm0.05)\times10^{5}~\mathrm{fm^{-1/2}},
\end{align}
corresponding to two-level $R$-matrix calculations that include
the interference between the subthreshold $2^{+}$ state and the
resonance at $E_R=4.35$~MeV, with updated constraints on the
$\alpha$- and $\gamma$-width amplitudes.
Each Gaussian component is assigned equal integrated weight, ensuring
that both physically admissible solutions contribute equally to the
posterior sampling.

The ground-state ANC $C_{0}$ retains the weighted flat prior
defined above, allowing the combined $(C_{0},C_{2})$ prior
to reflect both theoretical and experimental uncertainties.

\subsection{Posterior characteristics under the two-Gaussian ANC prior}

We now consider the posterior obtained with a two-Gaussian prior on the ANCs, 
which
encodes independent experimental constraints on the ANC $C_{2}$ of the
subthreshold $2^{+}$ state. This prior admits two physically distinct parameter 
regions:
a lower-$C_{2}$ and a higher-$C_{2}$. Each region yields its own posterior 
component in
$S_{E2}$, and the combined result is their equal-weight 
mixture.

\begin{figure}[H]
\centering
\includegraphics[width=0.48\textwidth]{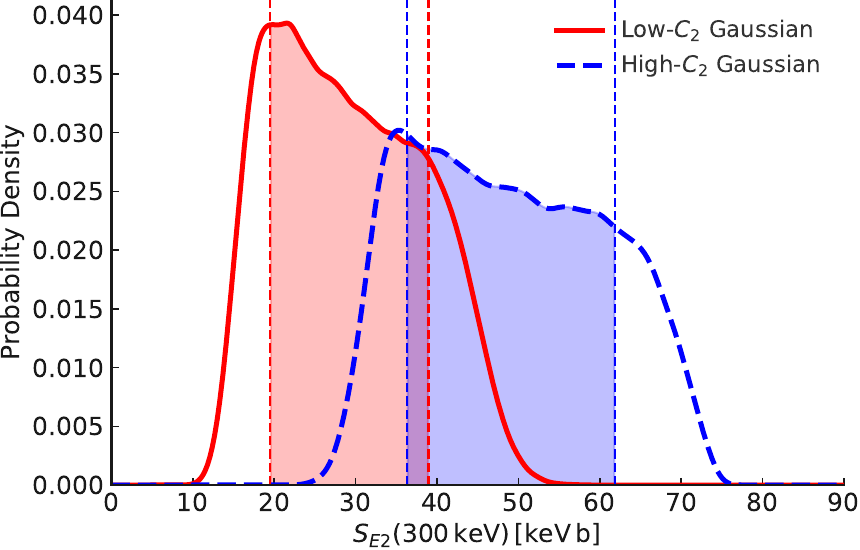}
\caption{Posterior distributions of $S_{E2}(300~\mathrm{keV})$ under the two-
Gaussian ANC prior. The solid (dashed) curve corresponds to the low-$C_{2}$ 
(high-$C_{2}$) component.  Relative to the flat-prior posterior, both 
distributions are narrower and more symmetric, reflecting the suppression of the
 small-$C_{0}$ regime that previously generated the high-$S_{E2}$ tail.}
\label{fig:SE2_2Gauss}
\end{figure}

\begin{table}[H]
\caption{Posterior summary for $S_{E2}(300~\mathrm{keV})$ under the two-Gaussian
 prior.
Credible intervals (CIs) are central: $68\% \equiv (16,84)$th and
$95\% \equiv (2.5,97.5)$th percentiles.}
\begin{ruledtabular}
\begin{tabular}{lccc}
 & Median & MAP & 68\% CI \quad 95\% CI \\
\hline
Low-$C_{2}$ & 28.35 & 19.9 & (19.12,\;39.61)\quad(14.84,\;46.36) \\
High-$C_{2}$ & 48.29 & 35.4 & (35.85,\;62.59)\quad(30.34,\;70.17) \\
Combined (50/50) & 37.68 & 34.7 & (23.25,\;55.60)\quad(15.96,\;68.30) \\
\end{tabular}
\end{ruledtabular}
\label{tab:SE2_2Gauss_summary}
\end{table}

Compared to the mixed prior described in Section \ref{quasiflat}, this distribution is significantly  narrower and  exhibits a reduced high-$S_{E2}$ tail.  The physical reason is that 
large-$C_{0}$
 (which suppress $S_{E2}$ by strong destructive interference) and very
small-$C_{0}$ configurations (which allow  large $S_{E2}$) are both constrained
by the two-Gaussian prior.  The resulting posterior reflects the set of 
$(C_{2},C_{0})$ combinations along the
interference corridor that are both supported by the data and consistent with 
the
independent constraints encoded in the two-Gaussian prior.

Note that the low–Gaussian prior, based on the value of $C_{2}$ determined from 
sub–Coulomb transfer \cite{Brune,Avila,Hebborn}, remains correlated with the ${}^{6}\mathrm{Li}$ ANC used  to normalize those measurements.

\section{Interference Geometry of the $E1$ and $E2$ Contributions}
\label{sec:ratio_geometry}

At $E=300$~keV the $E1$ strength is determined by the interference between the
subthreshold $1^{-}$ state and the broad $1^{-}$ resonance at $E_x=9.585$~MeV, 
with
the direct $E1$ transition being negligible.  In contrast, $S_{E2}$ is set 
by
the interference between the amplitude of the subthreshold resonance $2^{+}$  (controlled by  $C_{2}$) and the external direct-capture amplitude (controlled by $C_{0}$), which produces a narrow interference corridor in $(C_{2},C_{0})$ space.
The relative importance of the $E1$ and $E2$ components of the
${}^{12}\mathrm{C}(\alpha,\gamma){}^{16}\mathrm{O}$ cross section at
$E=300$~keV is determined by the interplay of the above mentioned mechanisms.
Since both $S_{E1}$ and $S_{E2}$ depend on different combinations of ANCs,
their ratio provides a compact summary of how the reaction dynamics reorganize
across the $(C_{2},C_{0})$ domain at different $C_{1}$.

\begin{figure}[t]
  \centering
  \includegraphics[width=\columnwidth]{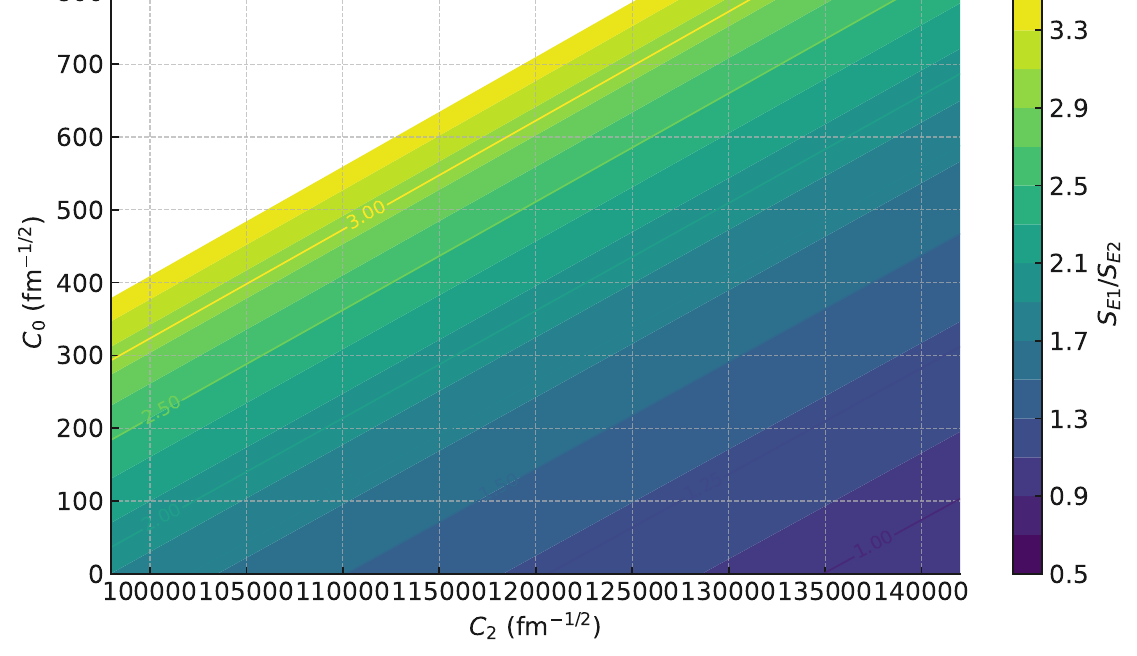}
  \caption{Ratio map $S_{E1}/S_{E2}$ at $E=300$ keV for the low-anchor
  $C_1=1.83\times10^{14}\,\mathrm{fm}^{-1/2}$ (yielding $S_{E1}=68.42$ keV\,b).
  Contours are numerically labeled; background shading helps visualize the
  interference corridor in $(C_2,C_0)$ space where the ratio varies slowly.
  This schematic complements the $E2$ interference surface by indicating regions
  where \(E1\) or \(E2\) dominates under the adopted ANC mapping.}
  \label{fig:SE1SE2ratio_low}
\end{figure}

\begin{figure}[t]
  \centering
  \includegraphics[width=\columnwidth]{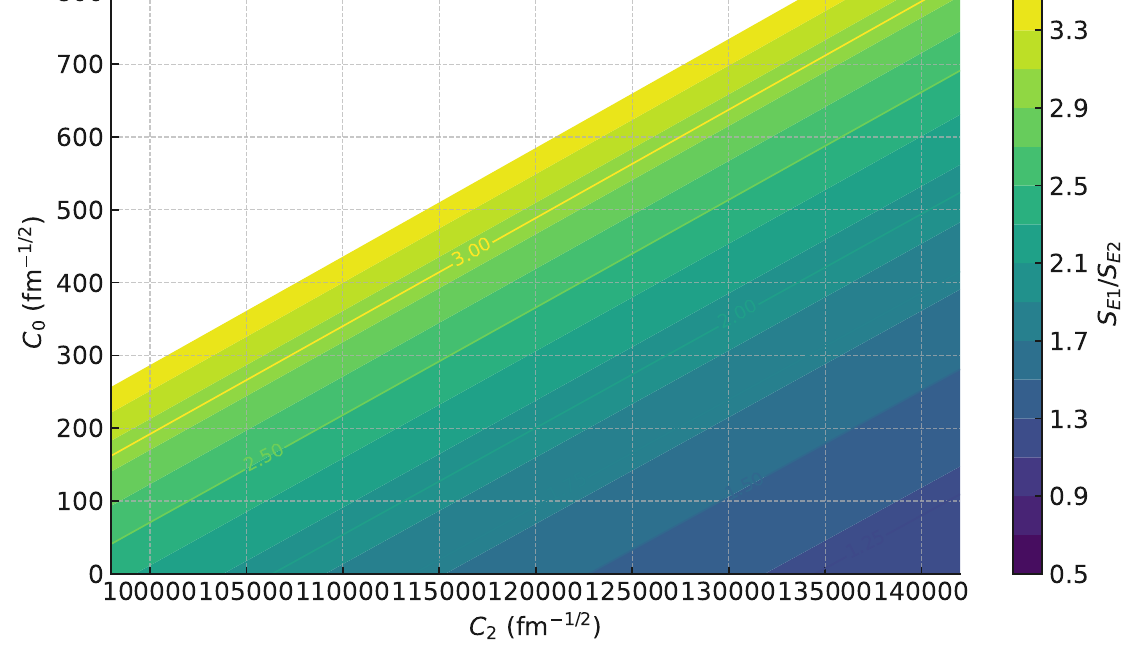}
  \caption{Ratio map $S_{E1}/S_{E2}$ at $E=300$ keV for the medium-anchor
  $C_1=2.08\times10^{14}\,\mathrm{fm}^{-1/2}$ (yielding $S_{E1}=85$ keV\,b).
  The rest of the caption is the same as in Fig. \ref{fig:SE1SE2ratio_low}.}
  \label{fig:SE1SE2ratio_mid}
\end{figure}

\begin{figure}[t]
  \centering
  \includegraphics[width=\columnwidth]{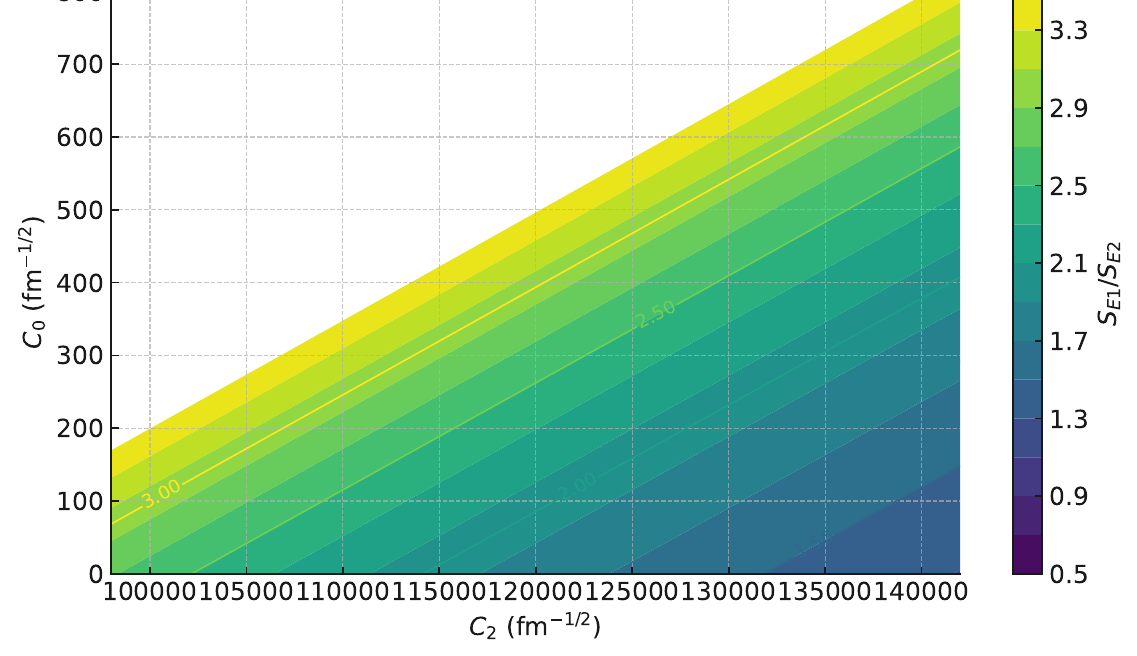}
  \caption{Ratio map $S_{E1}/S_{E2}$ at $E=300$ keV for the high-anchor
  $C_1=2.27\times10^{14}\,\mathrm{fm}^{-1/2}$ (yielding $S_{E1}=98$ keV\,b).
  The rest of the caption is the same as in Fig. \ref{fig:SE1SE2ratio_low}.}
   \label{fig:SE1SE2ratio_high}
\end{figure}

To visualize how these mechanisms compete, we examine the ratio
$R(C_{2},C_{0}\,C_{1}) \equiv S_{E1}/S_{E2},$
using the calibrated maps for $S_{E1}$ and $S_{E2}$.  
Figures~\ref{fig:SE1SE2ratio_low}, 
\ref{fig:SE1SE2ratio_mid}  and  \ref{fig:SE1SE2ratio_high}
show $R$ across the $(C_{2},C_{0})$ grid for three different representative ANCs
 $C_{1}$, 
low, medium and high.
The labeled iso-contours of $R$ trace how variations in $C_{2}$ and $C_{0}$
shift the relative strength of $E1$ and $E2$ capture.

Along the interference corridor, where $S_{E2}$ varies slowly,
the ratio $R$ changes gradually as well.  In contrast, motion
transverse to this corridor produces rapid variation in $R$,
reflecting the strong $C_{0}$ dependence of the $E2$ interference term.
Regions with small $C_{0}$ (weak external capture) support comparatively
larger $S_{E2}$ and therefore smaller $R$, while increasing $C_{0}$
enhances the external part of the $E2$ amplitude and strengthens destructive
interference, suppressing $S_{E2}$ and increasing the ratio $R$.

Thus, the $E1/E2$ ratio map provides a complementary geometric perspective on the reaction mechanism.  
The $E1$ contribution is governed primarily by the ANC $C_{1}$ of the subthreshold $1^{-}$ state, whereas the $E2$ contribution reflects the balance between the subthreshold $2^{+}$ amplitude scaled by $C_{2}$ and the external direct-capture path scaled by $C_{0}$.  
This balance follows the interference corridor that threads through the $(C_{0},C_{2})$ plane.  
The Bayesian posteriors developed in the previous sections can therefore be interpreted in terms of how probability mass arranges itself along this corridor and how that distribution simultaneously influences both $S_{E1}$ and $S_{E2}$.

\section{Astrophysical Connection: $S(300)$ and Black--Hole Remnant Mass}
\label{sec:BHconnection}

The $^{12}\mathrm{C}(\alpha,\gamma)^{16}\mathrm{O}$ reaction controls the carbon
--to--oxygen ratio
established at the end of core He burning. This ratio sets the mass and 
composition
of the resulting CO core, which in turn governs the onset and severity of
electron--positron pair formation in very massive stars. For sufficiently large 
CO--core
masses, pair creation softens the equation of state and drives dynamical 
expansion,
leading either to (i) repeated pulsational mass ejection (pulsational pair--
instability; PPI)\footnote{
Pair instability occurs when the stellar core becomes hot enough for 
energetic photons to convert into electron--positron pairs, reducing 
radiation pressure support and triggering rapid contraction followed by 
explosive oxygen burning.} 
or (ii) full thermonuclear disruption (pair--instability supernova; PISN).  
Whether a black hole remnant forms, and with what mass, is therefore directly 
linked
to the core mass at the end of helium  burning.

The key monotonic relationship is that a lower total $S$ factor at $300$ keV yields a larger CO--core mass.
A reduced total $S_{E1}+S_{E2}$ decreases the efficiency of $\alpha$--capture 
into $^{16}$O, leaving a carbon--rich core that ignites later and grows more before collapse. Stellar--evolution calculations
\cite{Wang2024,Abbott,Sukhold2016,Farmer2020,Tang2020}
consistently find:
\begin{itemize}
\item For low $S \lesssim100~\mathrm{keV\,b}$: CO cores become more massive.
Massive stars with masses $\sim60$--$90\,M_\odot$ undergo
direct--collapse--like evolution, producing low--spin black holes with
$M_{\rm BH}\sim50$-$70\,M_\odot$.
\item For intermediate $100\lesssim S\lesssim140~\mathrm{keV\,b}$:
stars enter the PPI regime, expelling several solar masses prior to collapse,
yielding smaller black--hole remnants.
\item For high $S\gtrsim140~\mathrm{keV\,b}$:
the CO core reaches the PISN domain and is completely disrupted,
leaving no black--hole remnant.
\end{itemize}

Recent gravitational--wave detections of massive, low--spin black holes in the
$\sim50$--$70\,M_\odot$ range \cite{Wang2024,Abbott,Wang}
thus favor a comparatively low $S$ within the astrophysical
range permitted by nuclear data.  

\begin{figure}[H]
\centering
\includegraphics[width=0.95\linewidth]{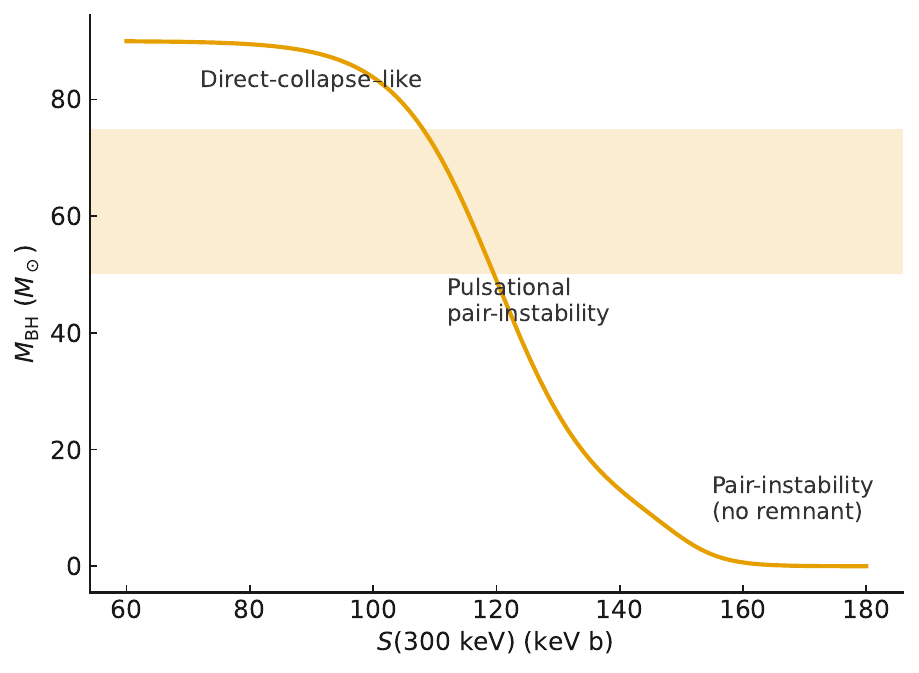}
\caption{Correlation between the total astrophysical factor
$S(300)=S_{E1}(300)+S_{E2}(300)$ and the expected black–hole remnant mass
from massive–star evolution. A lower $S(300)$ leads to a more carbon–rich core
and larger CO–core mass, favoring direct collapse and producing high–mass,
low–spin black holes ($M_{\rm BH}\sim 50$–$75\,M_{\odot}$). Intermediate
$S(300)$ values drive pulsational pair–instability (PPI), resulting in repeated
mass ejections and smaller remnants. Higher $S(300)$ pushes the core into the
pair–instability supernova (PISN) regime, in which the star is completely
disrupted and no black hole forms. The posterior range obtained in this work
($S(300)\approx 95$–$125$~keV\,b) lies squarely in the direct–collapse domain.}
\label{fig:BHvsS300}
\end{figure}

The posteriors obtained in Secs.~\ref{PosterSE1}--\ref{Priors}
show that the Gaussian prior centered on the lower experimental anchor for the
subthreshold $1^{-}$ ANC yields
\begin{align}
\mathrm{MAP}\; S_{E1} \approx 75.4~\mathrm{keV\,b},
\end{align}
while the Gaussian prior centered on the lower anchor for the subthreshold 
$2^{+}$ ANC
gives
\begin{align}
\mathrm{MAP}\; S_{E2} \approx 19.9~\mathrm{keV\,b}.
\end{align}
Together, these contributions correspond to
\begin{align}
S \approx 95.3~\mathrm{keV\,b},
\end{align}
placing the reaction rate squarely in the region associated in
Fig.~\ref{fig:BHvsS300} with direct--collapse–like black--hole formation
and remnant masses consistent with the observed $50$--$70\,M_{\odot}$ 
population.
Thus, the posterior regions identified in the present Bayesian analysis, particularly 
those corresponding to lower $C_{1}$ and higher $C_{0}$, fall naturally into this 
direct--collapse--like regime. Since the low values of $C_{1}$ and $C_{2}$ are determined from the sub–Coulomb 
transfer reactions \cite{Brune,Avila,Hebborn}, they are correlated with the 
${}^{6}\mathrm{Li}$ ANC used to normalize those measurements.

A key conclusion that emerges from this analysis is that the low values of the subthreshold ANCs $C_{1}$ and $C_{2}$ — obtained from the sub-Coulomb ${\alpha}$-transfer  \cite{Brune,Avila,Hebborn} — together with the high ground-state ANC $C_{0}$  supported by the latest measurements \cite{Gupta}, naturally produce total $S$-factors at $300$ keV fully consistent with the gravitational-wave observations of massive, low-spin black holes in the $50\,-\!70\,M_{\odot}$ range \cite{Wang2024,Wang}.

\section{Limitations of the fixed-parameter mapping.}

It should be noted that in the present analysis the $R$-matrix
parameters other than the ANCs were kept fixed at their reference
values. This approximation neglects possible correlations between the
ANCs and other $R$-matrix quantities, such as level energies, reduced
$\alpha$ widths, and interference phases. A fully self-consistent
treatment would require refitting all $R$-matrix parameters
simultaneously with varying ANCs using the same data set and
methodology as Ref.~\cite{deBoer}. Such a comprehensive
re-optimization, while more rigorous, is beyond the scope of the
present work. Nevertheless, it is expected that a full multi-parameter
refit would moderately reduce the apparent sensitivity of
$S_{E1}(300)$ and $S_{E2}(300)$ to the ANCs, without changing the
overall trends and conclusions presented here.
A complete $R$-matrix refit including ANC variations will be an
important subject for future work.

\section{Discussion}

The Bayesian posteriors obtained in this work show that the dominant 
uncertainties in the extrapolated 
${}^{12}\mathrm{C}(\alpha,\gamma){}^{16}\mathrm{O}$ 
astrophysical $S$–factors at $E=300$~keV arise from the 
subthreshold ANCs that determine the amplitudes and interference patterns of the 
capture pathways.

The $E1$ capture strength at $E=300$~keV is dominated by the subthreshold 
$1^{-}$ state at $E_{x}=7.12$~MeV.  
Quantitatively, this state contributes about $73$--$74\%$ of the total $E1$ 
amplitude at 300~keV, while the remaining $26$--$27\%$ arises from interference 
with the broad $1^{-}$ resonance at $E_{x}=9.59$~MeV.  
The role of the broad resonance is therefore chiefly to fix the interference 
phase and determine the detailed energy dependence, while the overall magnitude 
of $S_{E1}$ is controlled by the ANC $C_{1}$ of the subthreshold level.

Because the radiative $E1$ amplitude depends approximately quadratically on 
$C_{1}$, the posterior for $S_{E1}$ is strongly prior–sensitive.  
A flat prior over the accepted $C_{1}$ range yields a MAP near 
$S_{E1} \approx 95.5~\mathrm{keV\,b}$, while low– and high–Gaussian priors 
produce MAP values of $\approx 75.4$ and $\approx 93.2~\mathrm{keV\,b}$, 
respectively.  
Thus, the dominant source of uncertainty in the $E1$ contribution remains the 
subthreshold $1^{-}$ ANC, even when the interference with the broad $1^{-}$ 
resonance is taken fully into account.

In contrast, the $E2$ contribution is governed by coherent interference between 
capture through the subthreshold $2^{+}$ state (controlled by $C_{2}$) and the 
external direct capture to the ground state (controlled by $C_{0}$).  
The mapping $(C_{2},C_{0})\mapsto S_{E2}$ contains an extended interference 
corridor: motion \emph{along} this corridor keeps $S_{E2}$ nearly constant, 
while motion \emph{across} it changes $S_{E2}$ rapidly.  
Therefore, the shape of $P(S_{E2}\mid D)$ is determined by how the joint 
posterior $P(C_{2},C_{0}\mid D)$ distributes probability along the corridor.

With a flat prior on $C_{2}$ and a weighted prior that favors larger $C_{0}$, 
the posterior weight shifts toward regions where destructive interference is 
stronger, lowering $S_{E2}$ and truncating the low–$S_{E2}$ side of the 
distribution.  
However, the small–$C_{0}$ sector of the corridor remains allowed, producing a 
long high–$S_{E2}$ tail.  
The resulting $P(S_{E2}\mid D)$ is therefore \emph{right–skewed}: a well-defined 
peak at lower $S_{E2}$ with a probability tail extending toward higher values.

The two–Gaussian prior on $C_{2}$ suppresses both the very small–$C_{0}$ region 
that generates the high–$S_{E2}$ tail and the very large–$C_{0}$ region that 
drives $S_{E2}$ to low values.  
The posterior then separates into two narrow components associated with low– and 
high–$C_{2}$ solutions, and their equal-weight mixture yields the combined 
median
\begin{align}
S_{E2} \approx 38~\mathrm{keV\,b},
\end{align}
with significantly reduced tail structure.  
The narrowing of the posterior reflects the shorter contributing segment of the 
interference corridor, not a change in the underlying interference mechanism.

The credible intervals reported here therefore represent the present state of 
nuclear constraints on this reaction.  
Further reduction of uncertainty requires improved determinations of $C_{1}$ for 
the $E1$ channel and tighter constraints on $C_{2}$ and especially $C_{0}$ for 
the $E2$ channel.  Since the low values of $C_{1}$ and $C_{2}$ are determined from the sub–Coulomb 
transfer reactions\cite{Brune,Avila,Hebborn}, they are correlated with the 
${}^{6}\mathrm{Li}$ ANC used to normalize those measurements.

The astrophysical implications of the extracted $S$--factor are significant. 
The range of $S(300)$ favored by the ANC--constrained posteriors---particularly 
the combinations of lower $C_{1}$ and $C_{2}$ from sub-Coulomb transfer 
measurements \cite{Brune,Avila,Hebborn} and the larger ground-state ANC $C_{0}$ 
supported by recent high-precision determinations \cite{Gupta}---lies squarely 
in the regime where stellar-evolution models predict direct collapse to massive, 
low-spin black holes with gravitational masses of 
$\sim 50$--$70\,M_{\odot}$.  
Values near 
$C_{1}\approx (1.8\!-\!2.1)\times 10^{14}\,\mathrm{fm^{-1/2}}$ and 
$C_{0}\gtrsim 700\,\mathrm{fm^{-1/2}}$ naturally yield 
$S(300)\approx 90$--$120~\mathrm{keV\,b}$, precisely the range associated with 
direct collapse rather than pulsational or full pair--instability disruption.  
Thus, the gravitational-wave detections of massive, low-spin black holes provide 
an \emph{independent astrophysical preference for the low-ANC branch} of the 
subthreshold $1^{-}$ and $2^{+}$ states, reinforcing the central role that 
accurate ANCs play in constraining massive-remnant formation in nuclear 
astrophysics.

\section{Acknowledgments}
The author  thanks Alex Zhanov for technical assistance. The author also would 
like to acknowledge the assistance of  AI (OpenAI's  ChatGPT5) for editorial  
and formatting  support during  the preparation of the manuscript. 
 All scientific results, interpretations and  conclusions are solely those of 
the author. 

\appendix
\section{Marginal versus conditional probabilities}   
\label{MargCondProb}
In Bayesian inference it is important to distinguish between marginal and 
conditional 
probability densities:  
\begin{itemize}  
\item {Conditional probability density:}  
Describes the distribution of one parameter given that another is fixed.  
For example, $P(S \mid C)$ denotes the distribution of the $S$--factor 
when the ANC $C$ is assumed to have a specific value.  
\item {Marginal probability density:}  
Obtained by integrating (``summing over'') other variables, thereby removing 
them.  
For example,
\begin{align}
P(S) \;=\; \int P(S \mid C)\,P(C)\,dC
\label{eq:PS}
\end{align}
is the marginal density of the $S$--factor, averaged over all possible $C$ 
values 
weighted by their prior or posterior density.  
\end{itemize}
In short:\\
  Conditional: distribution of one parameter given a fixed value of another 
(e.g. $P(S\mid C)$). \\
Marginal: distribution after integrating out the uncertainty in the other 
variables (e.g. $P(S)=\int P(S\mid C)P(C)\,dC)$.

\section{Likelihood versus probability}  
\label{LiPr}
It is also crucial to separate the concepts of probability and likelihood:  
\begin{itemize}  
\item {Probability (density):}  
Quantifies how likely a future or hypothetical observation is, given a fixed 
model and 
parameters. For example,
\begin{align}
P(D\mid C)
\label{eq:PDC} 
\end{align} 
is the normalized probability of observing the data $D$ if the ANC $C$ has a 
particular value.  
 
\item {Likelihood:}  
The same expression viewed as a function of the parameter with data $D$ fixed:
\begin{align}
\mathcal{L}(C\mid D) \propto P(D\mid C).
\label{eq:LCD}
\end{align}
The likelihood is not normalized in $C$, but it provides the relative weight of 
different candidate values given the observed data.  
\end{itemize}  
In short: probability treats the parameter as fixed and the data as variable, 
while likelihood treats the data as fixed and the parameter as variable. 

For example,  the probability density of the data ${\tilde C}$ given the ANC  
$C$ is modeled by a Gaussian,
\begin{equation}
P(D\mid C) \;\propto\; 
\exp\,\left[-\frac{(\tilde{C}-C)^{2}}{2\sigma^{2}}\right].
\end{equation}
Viewed as a function of $C$, this expression is the \emph{likelihood},
\begin{equation}
\mathcal{L}(C\mid D) \;\propto\;
\exp\,\left[-\frac{(\tilde{C}-C)^{2}}{2\sigma^{2}}\right].
\end{equation}
The symbol ``$\propto$'' indicates proportionality: the right-hand side gives
the shape of the function in $C$, while the omitted normalization constant
ensures that $P(D\mid C)$ integrates to one over all possible data.  
In Bayesian inference, this normalization is irrelevant because $D$ is fixed
(the data have already been observed). What matters is the relative weight of
different $C$ values, which is fully specified by the likelihood.

\section{Marginal probabilities}
\label{app:marginals}

In Bayesian inference two types of marginal probabilities appear naturally.  
\begin{itemize}
\item {Evidence as a marginal over parameters.}  
The normalization factor in Bayes’ theorem,
\begin{equation}
P(D) \;=\; \int P(D\mid C)\,P(C)\,dC ,
\end{equation}
is the marginal probability of the data $D$, obtained by integrating 
over all possible parameter values $C$. This quantity, also called the 
\emph{evidence}, ensures that the posterior $P(C\mid D)$ is normalized.

\item {Prior as a marginal over nuisance parameters.}  
When a model involves several parameters, e.g.\ $(C_{0},C_{1},C_{2})$, 
one often specifies a joint prior $P(C_{0},C_{1},C_{2})$.  
The prior for a single parameter, say $C_{1}$, is then obtained by 
integration:
\begin{equation}
P(C_{1}) \;=\; \iint P(C_{0},C_{1},C_{2})\,dC_{0}\,dC_{2}.
\end{equation}
Thus $P(C)$ in Bayes’ theorem is itself a marginal probability distribution 
defined before the data are taken into account.
\end{itemize}

In summary, $P(D)$ is a marginal probability in \emph{data space}, while 
$P(C)$ is a marginal probability in \emph{parameter space}.

\section{Credible intervals.}
In Bayesian inference, two types of credible intervals (CIs) are commonly 
reported:
\begin{itemize}
\item {Central (CIs):} 
defined as symmetric in probability mass around the posterior median. 
That is, equal probability is excluded from both tails of the distribution 
(e.g. $16\%$  from each tail for a  $68\%$ central CI). 
Note that for skewed posteriors, the resulting interval is not symmetric 
in the parameter axis and may not contain the MAP.
\item {Highest posterior density intervals (HPDIs):} 
defined as the shortest interval(s) containing the specified probability mass. 
HPDIs always include the MAP and can be asymmetric.
\end{itemize}
The reported intervals in this work are the central CIs.

\section{MAP, median, and mean.}
\label{MAPMED}
The MAP (maximum a posteriori) is  the point of highest  probability density.  
The \emph{median} is the value that divides the posterior into two equal 
probability halves.  
The \emph{mean} is the expectation value, obtained by averaging the 
posterior.  
For symmetric posteriors all three coincide, but for skewed posteriors 
they generally differ.

\section{Kernel density estimation (KDE)}
\label{KDE}
When computing posteriors for $S_{E1}$ and $S_{E2}$, we generate large 
Monte Carlo samples from the ANC distributions and map them through the 
$R$--matrix formulas. A raw histogram of these samples would appear noisy and 
bin--size dependent. To provide smooth curves we use \emph{kernel density 
estimation} (KDE).

In KDE, each sample point is replaced by a small Gaussian ``bump.'' Adding all 
these bumps 
produces a smooth approximation to the underlying density:
\[
\hat{p}(x) \;=\; \frac{1}{Nh}\sum_{i=1}^{N}
K\,\left(\frac{x-x_{i}}{h}\right),
\]
where $K$ is the Gaussian kernel and $h$ is the bandwidth.  

In our figures KDE is used purely for visualization. The shaded $68\%$  bands 
are 
computed directly from the Monte Carlo samples, while the KDE curves guide the 
eye 
by showing the posterior shape more clearly.

\end{document}